\documentstyle[psfig,eqsecnum,aps]{revtex}

\begin{document}

\title{Kinetics
of coherent order-disorder transition in $Al_3 Zr$}

\author{ Laurent Proville \footnote{\small Present address: Groupe de Physique des Solides, UMR 7588 - CNRS, 
Universit\'es Paris 6 et Paris 7, 2 pl. Jussieu 75251, Paris Cedex 05, France}
 and  Alphonse Finel} 
\address{Laboratoire d'Etude des Microstructures\\
 ONERA - CNRS, BP72, 92322 Ch\^atillon Cedex, France}

\date{\today}

\maketitle

\begin{abstract}
 Within a phase field approach which takes  the strain-induced
 elasticity into account, the kinetics of the coherent order-disorder 
 transition is investigated
 for the specific case of  $Al_3 Zr$ alloy.
 It is shown that a microstructure with cubic $L1_2$ precipitates appears as 
 a transient state 
 during the decomposition 
 of a homogeneous disordered solid solution into a microstructure 
 with tetragonal $DO_{23}$ 
 precipitates embedded into a
 disordered matrix.
 At low enough temperature, favored by a weak internal stress,  only
 $L1_2$ precipitates grow in 
 the transient microstructure preceding  
 nucleation of the $DO_{23}$
 precipitates that occurs exclusively at the interface of the solid solution 
 with the $L1_2$ precipitates. 
 
 Analysis of  microstructures at nanoscopic scale shows
 a characteristic rod shape for the 
 $DO_{23}$ precipitates due to the combination of their 
 tetragonal symmetry and their  large internal stress.

\end{abstract}

\pacs{ 61.66.Dk, 68.35.Rh ,64.60.Cn ,64.60.My  }

\section{Introduction}

Macroscopic properties of materials depend to a large extend
on the microstructures they present at a mesoscopic scale.
For the case of alloys the most efficient way to control the formation of
microstructures is by phase transformations.
Well known prototypes are the $\gamma-\gamma'$ superalloys, as $Ni-Al$ 
(see \cite{Ardell}).
The microstructure of such binary alloys consists of stable ordered domains,
dispersed in a disordered fcc matrix. The symmetry of 
the ordered domains is called $L1_2$: the atoms of the minority species are 
placed at the corner of the fcc motif and the atoms of the
majority species occupies the center of the faces of the same motif. 

The macroscopic properties are controlled by the 
size of precipitates, their spatial distribution, and their ability 
to resist to coarsening. In this context, we present a theoretical 
study of the dynamics of microstructures in an aluminum based alloy, 
namely $Al-Zr$.

In this system, for low enough concentration, 
the equilibrium ordered
structure is the tetragonal $DO_{23}$ phase. Its motif 
is obtained from $L1_2$ structure
with anti-phase boundaries in (100) directions with the period of $3$. In  $Al-Zr$
the cubic $L1_2$ phase is known to be metastable at all temperature $T$. This is  confirmed by
ab-initio electronic calculations at $T=0 K$ \cite{Fontaine} and
\cite{Freeman} which
show that the energy difference
between $L1_2$ and $DO_{23}$ is about $0.863 \times  10^8 J/m^3$ (or $9.1 meV/atom$), 
in favor of $DO_{23}$.
However, the lattice misfit of $DO_{23}$ with respect to pure $Al$ is
significantly larger than $L1_2$. Hence, the interplay 
between the chemical energy and, for coherent microstructure, the elastic
energy may induce various precipitation processes.

The aim of the present work is to investigate the decomposition processes
and to analyze the resulting microstructures at mesoscopic scale.
In that aim, we use a phase field approach, where the incoherent
chemical energy is represented by a Ginzburg-Landau free energy
supplemented by a strain-induced elastic energy, in the form proposed
by  Khachaturyan \cite{turyan83,turyan93,Khacha00}. 
Using numerical simulations,
we show that elastic stress
favors the $L1_2$ order at low $Zr$ concentration and low temperature
and in that range the kinetics generates $L1_2$ rather than $DO_{23}$ precipitates.
If the coherent 
elastic interaction is not included, only $DO_{23}$ precipitates nucleate.

Our simulations prove that one may obtain metastable $L1_2$ phase
in specific conditions that correspond to the experimental
conditions leading to a $L1_2$
microstructure. The lifetime of that microstructure
may be infinite compared to the simulation time,
provided the temperature is not too high.
Once the $L1_2$ precipitates have grown at sufficiently low temperature,
the system may be brought to a higher temperature, where no $L1_2$
would nucleate if the system had been quenched directly to that 
temperature. 
If the already developed $L1_2$ microstructure is aged again at 
higher temperature, then the $DO_{23}$ precipitation starts and $DO_{23}$
precipitates nucleate exclusively along the interfaces between $L1_2$
inclusions and the matrix.
Once $DO_{23}$ precipitates are formed, 
they grow and consume  $L1_2$ domains.
The latest transformation cannot be reversed with decreasing temperature.

In Sec.\ref{sec2}, 
the general principles of the kinetic model are presented. In  Sec.\ref{sec3},
we outline how the microelasticity contribution can be included into the model. The implementation
of the phase transition kinetics, the results obtained and their interpretation are given
in Sec.\ref{sec4}. Conclusions and perspectives are drawn in Sec.\ref{sec5}.

\section{Ginzburg Landau Functional}
\label{sec2}

Our mesoscopic method is a time-dependent Ginzburg-Landau kinetics driven by a
functional with two parts: first a 
Ginzburg-Landau functional that includes the 
chemical interactions and the interface effects 
and second a strain-induced elastic energy.

The explicit form of the Ginzburg-Landau (GL) free energy is imposed
by symmetry rules. The first step consists of identifying 
the long range order (LRO) 
parameters that represent the ordered phases we want to study.
In the present situation, we should introduce the LRO parameters of $DO_{23}$ and $L1_2$. 
Nevertheless, to simplify the writing of the GL free energy,
we choose to replace the $DO_{23}$ LRO parameters with the $DO_{22}$ ones
as these phases are both tetragonal. What differs between them is only the periodicity 
of the anti-phase boundaries that is $2$ for  $DO_{22}$ instead of $3$ for $DO_{23}$.
As our aim is not to investigate the competition between the latter phases, our description
only requires to capture the tetragonality of the possible stable ordered phase.

The $L1_2$ phase is simple: it consists in three independent
parameters $\eta_1$, $\eta_2$, $\eta_3$ which
correspond to the amplitudes of the three waves that contribute to $L1_2$.
The waves correspond to the three vectors of the reciprocal space
$K_1 =(100)$ , $K_2 =(010)$, $K_3 =(001)$, respectively.
If the microstructure consists only in $L1_2$ domains embedded into a 
disordered fcc matrix,
the local concentration $c(R)$ would be given by

\begin{eqnarray}
c(R)=c_0(R)+ \sum_{j=\{1,2,3\}} 
{\eta_j}(R) \exp{(i 2 \pi  K_j. R)} 
\end{eqnarray}
where the quantities $c_0(R)$ and  ${\eta_j}(R)$ vary slowly in space.
For a $DO_{22}$ structure, the probability $c(R)$ 
to find a atom $Zr$ at position $R$
can be written as follows:

\begin{eqnarray}
c(R)=c_0(R)+ \sum_{j=\{1,2,3\}} 
{\eta_j}(R) \exp{(i 2 \pi K_j. R)} &+& 
\gamma_j(R) \exp{(i 2 \pi Q_j . R)}+\nonumber\\
&& \gamma_j^*(R) 
\exp{(-i 2 \pi Q_j . R)}
\end{eqnarray}
where $\gamma_j^*$ is the complex conjugate of the amplitude parameter $\gamma_j$
and $Q_1 = (1/2\ 1\ 0)$, $Q_2=(0\ 1/2\ 1)$ and 
$Q_3=(1\ 0\ 1/2)$ blong to the reciprocal cubic lattice.
This choice is not unique but is sufficient. For example, 
$(1/2\ 0\ 1)$ can be replaced by $(1/2\ 1\ 0)$);
each index $j$ correspond to a possible orientation of the tetragonal
transformation yielding $DO_{22}$.
For each of the three orientational variants there are
four translational variants. For example, for a perfect
$DO_{22}$ phase with the orientation $j=1$, 
one has four translational variants defined by
either $\gamma_1^*= \eta_1$, $\gamma_1^*= -\eta_1$
$\gamma_1^*=-i \eta_1$ $ \gamma_1^*= i\eta_1$. 
For simplicity we
drop the complex variants. It does not affect the
description as there are still two translation variants
for each orientation of the $DO_{22}$ phase.

We note that the state $\gamma_j=0$ and $\eta_j \neq 0$
for $j=1,2,3$ leads to a density $c(R)$  that describes a $L1_2$ phase.
A perfect $L1_2$ implies that the three (100)-type waves have the same amplitude.
As $L1_2$ preserves the cubic symmetry there is no orientational variant
but only four translational variants.

We now develop a uniform Landau functional as a function 
of $\{c_0,\eta_j,\gamma_j \}$.
The terms of this analytical function  $f_L(c_0,\eta,\gamma)=F_0 . {\hat f_L}$
are selected to fit with the symmetry of the fcc lattice, that is each term 
must be invariant under any operation of the fcc symmetry group.
Formally there is no other rule to realize a polynomial expansion 
but  the simplest form is probably the best. 
In practice, other conditions (see below) must be satisfied. We propose the following adimentional functional:

\begin{eqnarray}
{\hat f_L}&=& A_n (c_0-c_1)^n+\frac{A_2}{2} (c_2-c_0) \sum_i \eta_i^2
-A_3 \eta_1 \eta_2\eta_3
 \nonumber\\
&&
+\frac{A_4}{4} \eta_i^4 
+{K_3} \gamma_i^{2}\eta_i+\frac{K_4}{2} \sum_{j=i \pm 1} 
 \gamma_{i}^2 \eta_{j}^2\nonumber\\
&& +\frac{B_2}{2} (c_2-c_0) \gamma_i^2 
- \frac{B_4}{4} \gamma_i^{4} + 
\frac{B_{4}'}{2} \sum_{j=i \pm 1} \gamma_i^{2}\gamma_{j}^2
+\frac{B_6}{6} \gamma_i^{6} \label{func2}\\
\nonumber
\end{eqnarray}
where order parameters indices are written modulo 3. 
The hat symbol points out the adimensional quantities: here
${\hat f_L}$ is an adimensional free energy density.
The sets of coefficients $\{A_j \}$, $\{B_j \}$, $\{K_j \}$ and  $\{c_1, c_2 \}$
should be function of the temperature but 
are chosen as constant parameters for simplicity.
The first term in the right hand side of Eq.(\ref{func2}) 
is the disorder contribution as it does not depend on 
the order parameters. The power $n$ of that term may have two values $n=n_+=2$ and $n=n_-=8$
depending whether $c_0> c_1$ or not. 
We introduce that property to adjust
the topology of the Landau functional with
both experimental and ab-initio measurements as
described below. The continuity of the
first and second derivatives of the
Landau functional are preserved 
which is sufficient for the present case.

The next three terms with $A_2,A_3,A_4$ amplitudes are 
the contribution of the star $(1\ 0 \ 0)$. 
The four last terms associated with $B_2,B_{4},B_{4}',B_6$ are the contribution of
the concentration waves $\{Q_i\}_i$.
The expansion to the sixth order is required to obtain the linear stability
of both $DO_{22}$ and $L1_2$ phases for the same range of concentration.
The $\{K_4\}$ and $\{B_{4}'\}$ coefficients couple 
concentration waves with different orientations. 
They control the amplitude
of the potential  barrier  between the minima that correspond to ordered phases.
The $K_3$ term is the amplitude of the coupling between the waves
that belong to the same orientational $DO_{22}$ variant, $(K_i ,Q_i)$  and equivalents.

In fact, the precise form of the Landau functional term by term
has not a direct
influence on the mesoscopic microstructure providing that the functional is globally invariant 
with respect to the space group of the fcc lattice. The
important ingredients are the excess of free-energy associated with
interfaces and long-range elastic interactions between domains. 
Therefore, we keep the lowest order coupling term between the LRO parameters
associated with $(K_i ,Q_i)$ , i. e. , terms of the form $\gamma_i^2 \eta_i$.

The energy scale is fixed by the $F_0$ coefficient.
The parameters of the adimensional free-energy density
${\hat f_L}(c_0,\eta,\gamma)$ are adjusted
to fit the required qualitative 
thermodynamical properties. In
Fig.1 is plotted versus concentration $c_0$
the free energy  density $F_0.{\hat f_L}$ minimized with respect to the LRO parameters. The 
three types of minima correspond
to the disordered phase, the $DO_{23}$ (or $DO_{22}$) and $L1_2$ structures.
The common tangent constructions determine the regions
where ordered phases may coexist with the solid solution.
The concentrations 
$c_{DO_{23},a }$ and $c_{ DO_{23},b}$ are the limit of the region
where coexist both disordered solid solution with the $DO_{23}$ phase.
Similar quantities can be determine for the coexistence of solid solution with $L1_2$ phase: 
$c_{L1_2,a }$ and $c_{L1_2 ,b}$.
The  concentration $c_{L1_2,a}$ and $c_{DO_{23},a}$ are the solubility limit of the $L1_2$ and $DO_{23}$
phases, respectively.

In a non-uniform system, the order parameters $\eta_i$ and $\gamma_i$ and the local concentration $c_0$
are spatially dependant. Within a phase-field approach, 
these parameters vary continuously thought the system. The energy excess due to the
interfaces of precipitate, is expressed as a continuous Hamiltonian of the parameter
fields and their first derivatives. 
To the lowest order, this leads to the following Ginzburg-Landau (GL) free energy
density: 

\begin{equation}
f_{GL} =  F_0. [{\hat f_L}(c_0,\eta,\gamma)+  
\{\lambda_{c} |\nabla c_0|^2  + \sum_{i} \lambda_{\eta} 
\nabla^2 \eta_i
+\lambda_{\gamma} \nabla^2 \gamma_i \}] \label{GL}
\end{equation}
where the $\lambda$ coefficients are the weight of the gradient terms.
Total energy is given by $F_{GL}=\int f_{GL} dV$.
For the numerical implementation, we introduce a discrete space which is
a cubic sub-lattice with unit cell of linear size $d$. This length must be 
large enough to justify our
continuous approach and define the scale of one pixel in our simulations.
The total free energy can be expressed as a discrete sum
$F_{GL}=d^3 \sum_{L} {f}_{GL} $ where $L$ represents the set of the 
sub-lattice nodes.

We now describe the physical requirement we use 
to adjust the Ginzburg-Landau functional.
First,
$c_{L1_2 a}$ and
$c_{DO_{23} a}$ (see Fig.1) 
are the solubility limit of both $L1_2$ and $DO_{23}$ if each
phase is supposed to coexist alone with solid solution. 
The  $c_{DO_{22} a}$ is given by the 
uncoherent 
phase diagram and $c_{L1_2 a}$
has been estimated by the measure of the lattice parameter of the solid solution
by X ray using Vegard's law \cite{Zedalis-Fine-86}. 
For $Zr$ at temperature $T=425^o C$,
$c_{DO_{22} a}=0.0308$  and  
$c_{L1_2 a}=0.0426$ at.$\%$ \cite{Zedalis-Fine-86}.

Second,  to analyze the competition between the metastable $L1_2$ and stable
$DO_{23}$ structures during the precipitation and aging processes, the free energy difference between
both phases is an important quantity. In order to estimate this difference
we refer to the theoretical
studies of the formation energies obtained with ab-initio electronic structure
methods \cite{Fontaine,Freeman,Meschter,Watson}. In \cite{Fontaine}, 
it is found that ground-state of $Al_3Zr$ is indeed
$DO_{23}$, and its structure is stabilized by the relaxation of the atomic positions inside
the elementary motif. 
The difference of energy between  $L1_2$ and 
$DO_{23}$ is found to be $\delta_{L1_2}=0.863\times 10^8 J/m^3$. In term of a phase field approach,
this corresponds to an uncoherent energy at zero temperature.
We assume that for low enough temperature, the free energy difference $\delta_{L1_2}$
do not vary strongly. Therefore, the ab-initio quantity  $\delta_{L1_2}$
is used to fix the scale $F_0$ of the free energy (Eq.\ref{GL}) through the relation
$F_0= {\delta}_{L1_2}/ {\hat \delta_{L1_2}}$ where ${\hat \delta_{L1_2}}$ is the 
energy difference between the corresponding $L1_2$ and 
$DO_{23}$ minima of the adimentional free energy $\hat f_L$ which
has been minimized taking into account the conservation of the $c_0$ concentration parameter that yields
the common tangent construction (see Fig.1).

Finally, another very important feature of the GL functional is the excess of energy of the
interface between solid solution and precipitates, noted $I_o$ with $o=\{L1_2, DO_{23} \}$. 
These quantities play a role in nucleation and growth process.
The interface energy of ordered precipitates in the solid solution
can be measured experimentally at very low supersaturation.
The Lifshitz-Slyozov-Wagner theory \cite{LSW} gives the interface energy 
as a function of the diffusion coefficient which is 
physically measured. 
Unfortunately the few we found in literature (see \cite{Zedalis-Fine-86}) about such a measurement
is not satisfactory
as the interface energy measured for $L1_2$ precipitates $I_{L1_2}$
is hundred times larger than the usual values. Thus
we choose to estimate $I_{L1_2}$ to 
a value of $10 mJ/m^2$ which is the order of magnitude of interface
energies measured in aluminum compounds. 
As we did not find measurement in literature concerning interface
energy of $DO_{23}$ precipitates $I_{DO_{23}}$, 
we choose for $I_{DO_{23}}$ a similar value to that of  $I_{L1_2}$
because there is no physical reason these two quantities to differ by an order of magnitude. 
We adjust the interface energies of the ordered phases $\{o\}$ of
the adimensional GL functional ${\hat I}_o$
such as $F_0 d {\hat I}_o=I_o$.
It implies ${\hat I}_{DO_{23}}/{\hat I}_{L1_2}= {I}_{DO_{23}}/{ I}_{L1_2}$.
With some difficulties,
we managed to adjust
 the GL functional such that both
${\hat I_{DO_{23}}}$ and ${\hat I_{L1_2}}$  have the same order of magnitude.
The $F_0$ being fixed by the second criteriion stated previously,
it imposes $d= I_o/{\hat I}_o. {\hat \delta_{L1_2}}/{\delta}_{L1_2}$
so we can define the scale of our simulation.
In order to investigate the nanometer scale we choose 
$d = 1 nm$, then
the $\hat{f}$ functional must verify
$ {\hat I_o} / {\hat \delta}_{L1_2} \approx  0.26$.
The only way we found to satisfy the previous criteria is to introduce
the non-symmetric power $n_{\pm}$ 
for the term $(c-c_1)^{n_\pm}$ in the GL functional.
Finally 
we obtain 
$c_{DO_{22} a}=0.0308 \%$,  
$c_{L1_2 a}=0.5 \%$  and 
$F_0. d. {\hat I_{L1_2}}= 8 mJ/m^2$ ,
$F_0. d. {\hat I_{DO_{22}}}=9 mJ/m^2$.
The pixel of our simulation represents a cube of size $d = 0.5 nm $.


\section{Microelasticity Contribution to Free Energy}
\label{sec3}

As described in \cite{Khacha96},
the elastic energy $E_{el}$ is calculated assumsing that the local strain
$(\epsilon_{kl})$ induces a relaxation 
that is calculated by setting a small volume $dV$ of the bulk
to the mechanical equilibrium.
It is supposed that the time needed to reach the mechanical 
equilibrium is negligible
compared to the typical diffusive time of the ordering process.
The key point of the phase field theory for alloys is that the stress free 
strain tensor can be expressed as a function of the local LRO
parameters and local concentration $c_0 (R)$.

The geometrical operation to transform a cubic unit cell of the solid solution 
with lattice parameter $ \overline a $
into the cubic unit cell of the $L1_2$ phase
with lattice parameter $ a_{L1_2} $ is given by  the tensor:

\begin{equation}
\epsilon^{L1_2}_{kl}= \delta_{kl} \
(a_{L1_2}-{\overline a})/{\overline a} 
\end{equation}
where $\delta_{kl}$ is the unity tensor.
The lattice parameter   $ a_{L1_2}$ has been measured for
a perfect $L1_2$ phase \cite{Zedalis-Fine-86} i. e. , with stoichiometry  $0.25$  at. $\%$ $Zr$ and 
$\overline a $ is extrapolated from the lattice parameter $a_0$ of pure aluminum $Al$,
using Vegard's law.

We note $ a_{DO_{23}}$ and $b_{DO_{23}}$ the lattice parameters of 
the tetragonal $DO_{23}$ phase at stoichiometry  $0.25$  at. $\%$ $Zr$.
The geometrical operation to transform a cubic unit cell of the solid solution 
into the tetragonal elementary cell of the $DO_{23}$ phase
is given by  the tensor $\epsilon_{kl}^{DO_{23}}(p)$ if
the orientational variant corresponds to the association of the $(K_p,Q_p)$ waves
with $p=1,2\ or\ 3$: the cell is dilatated in either $(100)$, $(010)$ or $(001)$ direction.
Here we choose as an example
the orientational variant associated to $(K_1,Q_1)$ and
the cubic unit cell is dilated in the direction $(100)$ so
as $b_{DO_{23}}>{\overline a} > a_{DO_{23}}$.
\begin{eqnarray}
\epsilon^{DO_{23}}(1)=   \frac{b_{DO_{23}}-{\overline a}}{{\overline a}}
\left( \begin{array}{ccc}
1 & 0 & 0 \\
0 & t & 0 \\
0 & 0 & t \\
\end{array} \right) \label{strainDO23}
 \end{eqnarray}
where $t=(a_{DO_{23}}-{\overline a})/(b_{DO_{23}}-{\overline a})$. 
Tensors $\epsilon_{kl}^{DO_{23}}(2)$ and $\epsilon_{kl}^{DO_{23}}(3)$
derive from $\epsilon_{kl}^{DO_{23}}(1)$ by permutation of the diagonal coefficients.
For the general expression of the local strain,
we propose the following form (see \cite{Khacha00}):
\begin{equation}
\epsilon_{kl}^{0}(R)=  \epsilon_{kl}^{00}(0)\ \psi_0(R) +
\sum_{p=1}^{3}  \epsilon_{kl}^{00}(p)\ \psi_p(R) \label{epsil}
\end{equation}
where
$\psi_p(R)=(\gamma_p(R))^2$, and $\psi_0(R) = (c_0(R) - \overline c)$.
The tensor coefficients $\epsilon_{kl}^{00}(p)$ are chosen in a such way that 
$\epsilon_{kl}^{0}(R)=\epsilon_{kl}^{L1_2}$ if 
a relaxed $L1_2$ inclusion is at the position  $R$ and 
$\epsilon_{kl}^{0}(R)=\epsilon_{kl}^{DO_{23}(p)}$ if a $DO_{23}$ inclusion
is in the same  position with the orientational variant $p$.
The Eq.\ref{epsil} is rewritten in a compact form: $ \epsilon^{0}(R)= 
\sum_{p=0}^{3}  \epsilon^{00}(p)\ \psi_p(R)$ where the $p$ indice varies now from $0$ to $3$.
The strain-induced elastic energy can be computed following the model presented in \cite{Khacha96}
which gives:
\begin{equation}
E_{el}=\frac{1}{2}
\sum_{p,q} \int [B_{pq}
 ] {\tilde \psi_p}^*{\tilde \psi_q} dK^3 \label{Eel}\end{equation}
where ${\tilde \psi_p}$ is the Fourier transform of the function
$\psi_p$. There
\[B_{p,q}(K)=\lambda_{ijkl}\ { \epsilon_{kl}^{00}(p)} \
{\epsilon_{ij}^{00}(q)}- b_{pq}(K)\]
and $b_{pq}(K)=\sigma_{i,j}^{00}(p)\ K_i \ G_{j k} \ K_l \ \sigma_{kl}^{00}(q)$ where
$\lambda_{ijkl}$ is the elastic tensor and
$\sigma_{ij}^{00}(p)=\lambda_{ijkl} \epsilon_{kl}^{00}(p)$. The tensor $G_{jk}$ is the elastic
Green function. The tensor $\lambda_{ijkl}$ is 
assumed to be homogeneous in space and
the simulations are realized with 
elastic coefficients of aluminum \cite{Lazarus}.

\section{Kinetics of the phase transformation}
\label{sec4}

The total energy is given by the sum $F=F_{GL}+E_{el}$ (Eqs.(\ref{GL}) and (\ref{Eel})). At mesoscopic scale,
the kinetics of the phase transition is well described by a phase field method (see \cite{Lebowitz}).
In the context of a phase field approach,
the local composition $c_0$ is a conservative order parameter and  
thus its time evolution is driven by the Cahn-Hilliard equation:
\begin{equation}
\frac{\partial c_0(R ,t)}{\partial t}=
L_c \bigtriangleup \frac{\delta F }{\delta c_0(R ,t)} + \upsilon^{c}(R,t) \label{KinC}
\end{equation}
and for the non-conservative LRO parameters the kinetic is given by 
\begin{equation}
\frac{\partial \eta_j }{\partial t}=-L_{\eta} 
\frac{\delta F}{\delta \eta_j } + \upsilon^{\eta_j}(R,t)\label{KinEta}
\end{equation}
\begin{equation}
\frac{\partial \gamma_j}{\partial t}=-L_{\gamma}
\frac{\delta F}{\delta \gamma_j} + \upsilon^{\gamma_j}(R,t)\label{KinGam}
\end{equation}

\noindent
where $\upsilon$'s are stochastic terms.
To simulate thermal fluctuations, it is useful to introduce the Langevin noise which consists in
assuming a white space-time noise for the stochastic terms and no cross-correlation between
each other.
Numerically, the random functions $\upsilon$'s are implemented with a gaussian probability
density \cite{NR}.


The set of Eqs.[(\ref{KinC})-(\ref{KinGam})]
is the so called Time-Dependant Ginzburg-Landau equation \cite{Khacha00}
and they can be derived from the microscopic \"Onsager equation with 
respect to the occupation probability of the solute atoms \cite{turyan83}. 
Numerically, for simplicity, we choose 
$L_{\gamma}=L_{\eta}=L_c=1$. So 
in order to deduce the approximative time unit of our simulations, our numerical 
time must be divided by the numerical value of the diffusion coefficient.
The Langevin noise provides a very primitive description of
the thermal fluctuations. All though it is the simplest and the less controversal way 
to implement the temperature in the simulation.

Equations [(\ref{KinC})-(\ref{KinGam})] are integrated on a system that is invariant along
$z$-axis to save computation time. The initial state of the system is a uniform
solid solution at concentration $c_{L1_2,a}<c_0={\overline c}< 0.25$ 
and all the set of LRO parameters  are set to zero value,
which represents an unstable disordered phase. The initial time of our simulation 
can be considered as an instantaneous quenching of the material.

To represent the microstructures we choose to color the $DO_{23}$
precipitates with either blue or red depending on their translational
variant. The four translational variants of $L1_2$ are allocated to four
other paler colors. The disordered phase is colored with black.
The gray scale is sufficient to distinguish the different types of precipitates if 
the translational variants are forgotten.
On Figs.[(\ref{fig4a})-(\ref{fig4c})], 
we present the dynamics of the phase transition from
disordered solid solution to a microstructure with ordered precipitates
embedded into a disordered matrix. The different sequences are realized for
different average compositions $\overline c$ and different temperatures $T$ .
\newpage

First, one notes the specific rod shape of the $DO_{23}$
precipitates. 
The tetragonal symmetry combined with a large misfit
$(b_{DO_{23}}-{\overline a})/{\overline a}$
which involves a large intrinsic stress is well known
to induce such a  pattern \cite{Khacha96}. 
On the very last picture of Fig.2
the facets of the $DO_{23}$ inclusions  
correspond to the habit planes with orientation
of around $20$ degrees with respect to the (1 0 0) directions.
For any couple of external variables, namely the
temperature  $T$ and the composition $ \overline c$ the late stage of the
kinetics is a microstructure which contains exclusively the $DO_{23}$
precipitates embedded in a disordered matrix. Nevertheless $L1_{2}$
structures may appear in the early regime of the dynamics (see Fig.3). As the $L1_{2}$ ordered precipitates involve a weak
misfit $(a_{L1_2}-{\overline a})/{\overline a}$ compared
to $(b_{DO_{23}}-{\overline a})/{\overline a}$, their shape is spherical. These spherical inclusions
nucleate only at low temperature $T$ 
and low $Zr$ concentration $\overline c$. At low enough
values of $T$ and for saturation 
$\overline c$ close enough from the solubility limit, 
the microstructure contains exclusively the
$L1_{2}$ precipitates inside the solid solution. Since the $L1_{2}$ phase
is metastable, any grain of this phase should not resist to thermal
fluctuations and thus no $L1_{2}$ precipitates should grow in the microstructure.
It is actually what is observed if the elastic energy $E_{el}$ is neglected
in our simulations. However, in the limit of low $Zr$ saturation, 
the kinetics of the transition drives the system to a  transient $L1_{2}$ microstructure
which is favored by its intrinsic stress which is weaker than that of $DO_{23}$. 
If the temperature $T$ is not too high, i. e., $T<T_{L1_2} \approx 500 K$ (see Fig.2),
the $L1_{2}$ precipitates can even grow by
consuming the very few solute atoms contained in the disordered matrix. 
This implies that the solid solution becomes poor in solute and
therefore the 
nucleation of $DO_{23}$ precipitates is no longer possible in the disoredered matrix.
In fact, if temperature is lower than $T_{L1_2}$, 
the $L1_{2}$ microstructure may survive for a  time much longer than the computation time. 
Nevertheless, 
a gradual increase of the temperature reveals 
 the process of
nucleation  of the $DO_{23}$ precipitates (see Fig.2). A remarkable result
is that nucleation of the stable phase occurs at the interface of
the $L1_2$ precipitates with the solid solution
where the local concentration $c_0$ is high enough. One note that
the  preferential growth  of  $DO_{23}$ precipitates
occurs at temperature higher than $T_{L1_2}$
that demonstrates the robusness of the metastable $L1_{2}$ microstructure with respect to 
thermal fluctuations.

Similar simulations at larger saturation $\overline c$
show that both the $DO_{23}$  and  $L1_{2}$ precipitates 
nucleate and may coexist in the microstructure (see Fig.4). 
In that range of $\overline c$, the $DO_{23}$ inclusions
can nucleate at places different from $L1_{2}$ precipitates
because the solute atoms have not been
consumed by the growth of $L1_{2}$ inclusions. 
Once saturation of the solid solution is locally dried up,
the $DO_{23}$ precipitates grow to the expense of the  $L1_{2}$ grains
via the solute diffusion throught the matrix. Then
the persistent $L1_{2}$ precipitates are
localized relatively far from the $DO_{23}$ grain. 
For a given temperature,
if the average concentration $\overline c$ is increased again,  the grains with 
different phases nucleate in neighboring regions of the supersaturated solid solution and  
the $DO_{23}$ inclusions absorbe the $Zr$ matter of $L1_2$ inclusions.

One remark on Figs.[(2)-(4)]
that the orientational variant of $DO_{23}$ which is combination of
waves $(K_3,Q_3)$ is inhibited. It is because  the  precipitates with such variant
cannot relax their elastic energy 
because of z-invariance. Qualitatively it is not a problem
as there are still two orientational variants for the $DO_{23}$ phase. 

\section{Conclusion and Perspective}
\label{sec5}

The present paper treats of the specific case of
the interplay between the $L1_2$ metastable phase and the $DO_{23}$ ordered
ground-state during the order-disorder transition in $Al_3 Zr$ alloy. It is proved that for a
sufficiently low temperature and weak solute saturation the metastable phase
nucleates before the stable phase. It is the result of the dynamics
of the phase transition that is deeply influenced by the microelasticity
induced by the strain of precipitates. The metastable ordered
phase with a weak internal stress may be favored with respect to the ordered
ground-state which induces a much larger strain.   
The strain-induced elasticity may play a role at the 
early regime of the dynamics of the phase transition.
Furthermore, depending on external variables, temperature and composition,
we found different kinetics for the vanishing 
of the  $L1_2$ precipitates.
At low temperature and low saturation of solute, the  $DO_{23}$ precipitates
grow preferentially at the interfaces of $L1_2$ inclusions with solid solution.
At higher temperature or equivalently higher $Zr$ saturation,
the  $DO_{23}$ precipitates nucleate into the solid solution.
Once $DO_{23}$ precipitates have nucleated, they grow at expense of the  $L1_2$ 
structure that disappears. Actually,
such phenomena have not yet been observed experimentally. 
With this respect, the phase field method can be considered  as a predictive method, though
experimental confirmation is now required. On that trail,
microscope analysis are programmed in the Laboratoire d'Etude des Microstructures (ONERA).

It is well known experimentally that 
the  $L1_2$ microstructure has better mechanical properties than $DO_{23}$. 
As the $L1_2$ is metastable, the degradation
of the mechanical properties with increasing temperature cannot be avoid. Nevertheless our results
allow to hope that it is possible to increase the robusness of the $L1_2$ microstructure playing with 
elastic interaction.
We expect our study will contribute 
to the  understanding of precipitates formation and to improve the control of the alloys synthesis.
To that purpose,
the phase field method we use, can be extended for other alloys with similar
phase transition as $Ti_3 Al$ or $Pd_3 V$.

Finally our calculations are only valid for coherent sample at the nanometer
scale. It is possible to investigate non-coherent effect introducing
dislocations in the phase-field method as it is discribed in both \cite{Khatcha-2000} \cite{Alphonse-2000}. 
It is of great interest to perform such simulation in the case of $Al_{3}Zr $ 
where discontinuous precipitations and dislocations modify
strongly the precipitation process \cite{Nes,Ryum} and play an important role in
macroscopic properties of materials.

 
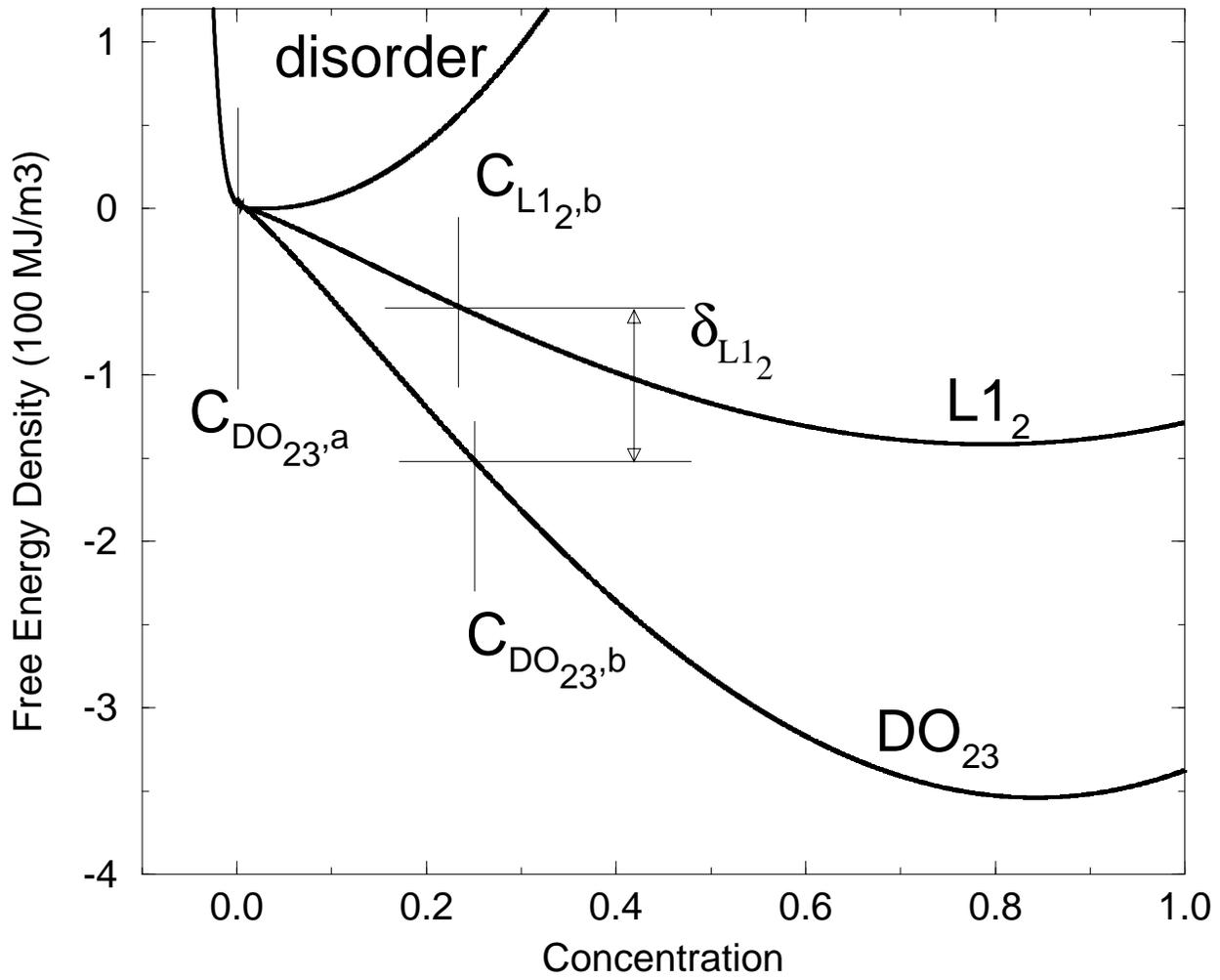
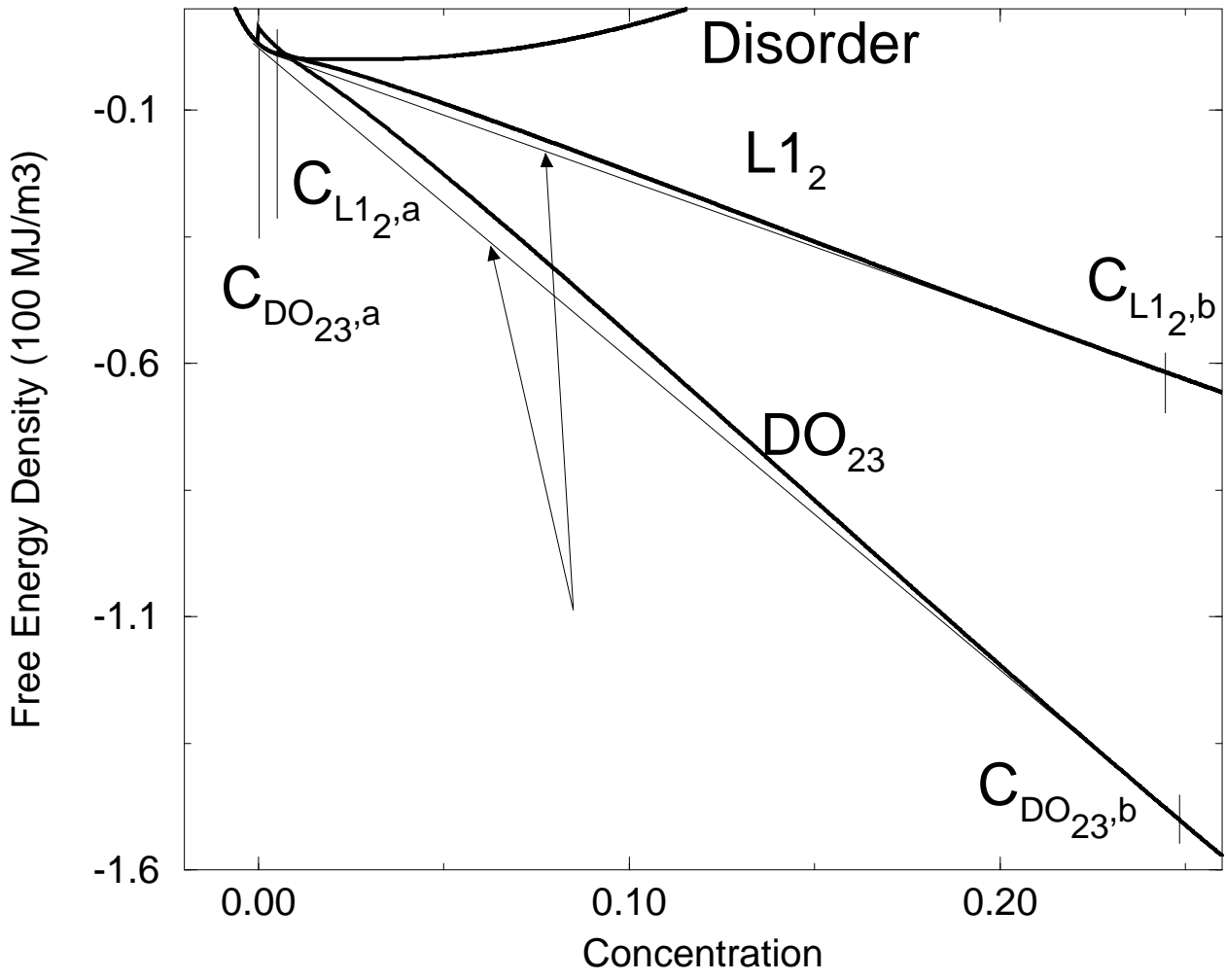
\end{document}